\documentclass[11pt]{article}
\usepackage[english]{babel}
\usepackage[utf8]{inputenc}
\usepackage[T1]{fontenc}
\usepackage{graphicx}
\usepackage{amsmath,mathrsfs,dsfont}
\usepackage[all]{xy}
\usepackage{graphics}
\usepackage{tensor}
\usepackage{verbatim}
\usepackage{nicefrac}
\usepackage{amsfonts}
\usepackage{amssymb,latexsym}
\usepackage{color}
\usepackage{mathrsfs,dsfont}
\usepackage{appendix}
\usepackage{slashed}
\usepackage{cite}
\usepackage{comment}
\DeclareMathOperator{\Tr}{Tr}
\newcommand{\nn}{\nonumber}
\DeclareMathOperator{\threeD}{\hspace{-2pt} {}^{(3)} \hspace{-2pt} \textit{D}}
\DeclareMathOperator{\threeR}{\hspace{1pt} {}^{(3)} \hspace{-2pt} \textit{R}}

\begin{document}

\title{IR Horava-Lifshitz gravity coupled to Lorentz violating matter:\\ A spectral action approach}
\author{D.~V.~Lopes\thanks{danielvieira1@gmail.com },\ \ A. Mamiya\thanks{arthurmamiya@gmail.com}\ \  and A. Pinzul\thanks{apinzul@unb.br}\\
\\
\emph{Universidade de Bras\'{\i}lia}\\
\emph{Instituto de F\'{\i}sica}\\
\emph{70910-900, Bras\'{\i}lia, DF, Brasil}\\
\emph{and}\\
\emph{{International Center of Condensed Matter Physics} }\\
\emph{C.P. 04667, Brasilia, DF, Brazil} \\
}
\date{}
\maketitle

\begin{abstract}
We continue our study of Horava-Lifshitz type theories using the methods of the spectral geometry. In this work we construct the infrared action of gravity and matter coupled to gravity in the most general way respecting the foliation preserving diffeomorphisms. This is done with the help of the spectral action principle based on some generalized Dirac operator. The gravity part reproduces the infrared limit of the Horava-Lifshitz gravity, while the matter part gives the generalization of the earlier suggested models. Due to the fact that the same Dirac operator is used in the construction of both sectors, the parameters of the gravity and matter parts are related. We expect that this potentially could naturally exclude fine tunings needed to get some desired properties as well as open new possibilities for the experimental tests of the model.
\end{abstract}

\section{Introduction}

In view of the ongoing search for the theory of Quantum Gravity and taking into account that neither String Theory nor Loop Quantum Gravity could not yet be considered as a completely satisfying solution, any attempt in this direction should be welcomed. One of the recent proposals is the so-called Horava-Lifshitz (HL) gravity \cite{Horava:2009uw}. This theory is based on the foliation-preserving diffeomorphisms (FPDiff) instead of the full diffeomorphisms of space-time. While the proposal has several interesting features expected from a consistent theory of quantum gravity, such as the power-counting renormalizability and ultraviolet spectral dimension of two \cite{Horava:2009if,Pinzul:2010ct,Carlip:2012md}, it suffers from several problems (for the detailed account on the HL gravity and some of its applications, see \cite{Padilla:2010ge,Sotiriou:2010wn,Mukohyama:2010xz}). One of the major drawbacks of the proposal, from our point of view, is a very large number of free parameters, which is of order of 100 \cite{Blas:2010hb}. To reduce this number one has to impose very restrictive conditions like the detailed balance condition \cite{Horava:2009uw}, which are neither natural nor satisfactory experimentally (see, though, \cite{Vernieri:2011aa}).

Another, somewhat related, point is the coupling of the gravity based on FPDiff to matter. There are some arguments \cite{Kimpton:2013zb} that in contrary to the gravity part the matter sector should respect the symmetry under full diffeomorphisms (or it should be extremely fine-tuned). Clearly this is very non-natural and one would like to have a more symmetric formulation of the theory where both, gravity and matter sectors, are based on FPDiff or one should give more fundamental reasons why matter should respect much larger symmetry.

In this paper, we are advocating the first possibility, i.e. that both sectors have FPDiffs as the fundamental symmetry. This is achieved via the spectral action principle \cite{Chamseddine:1996zu}. The main idea is that both sectors are controlled by the same object - some physically relevant Dirac operator. Due to this, both sectors are not independent and the parameters of the gravitational part are related to the parameters of the matter sector. Earlier this strategy was successfully applied to the Standard Model \cite{Chamseddine:2008zj}. We are applying this approach to construct the most general infrared (IR) action of the HL gravity coupled to fermionic matter. The action is based on some natural generalization of the standard Dirac operator. In our earlier works \cite{Pinzul:2010ct,Pinzul:2014mva} we used the same type of Dirac operator to analyze the spectral dimension of the HL space-time (which was confirmed to be 2 for an arbitrary curved space-time) and to study the geodesic motion of a test particle in the HL gravity. The main result of the analysis of the geodesic motion was that in the case of the non-minimal coupling it differs from the motion naively calculated from the underlying Riemannian metric. Of course this result was not a big surprise, but what is important that in our approach the deviation from the Riemannian geodesics is controlled by the same Dirac operator that is used in the construction of the HL modification of the Einstein-Hilbert action. Potentially this might lead to some cancellations of the experimentally problematic effects. This urgently requires further studies. In the current work, we take this proposal to the next level and study the most general IR action in both, gravity and matter, sectors. Here IR means energies much lower then some typical quantum gravitation scale where the higher order terms in the HL gravity become relevant. Of course, this includes the range accessible to the present experiments and well beyond. The main results of the paper are the following:

i) On the gravity side: The most general IR action of the HL type is derived as the spectral action for some generalized Dirac operator. The non-trivial result is that to obtain the most general form of the gravity action, one has to add P-violating terms to Dirac operator.

ii) On the matter side: Using the same Dirac operator as in (i) and the matter part of the spectral action, we obtain the most general Lorentz violating fermionic action based on FPDiff and consistent with the Standard Model Extension (SME) of \cite{Kostelecky:2003fs}. In our approach, the free parameters of SME are expressed in terms of the parameters of the Dirac operator. A very important result is that these parameters also depend on the geometry of the space-time, which can make Lorentz violating effects very small in a weak gravitational field.

iii) The combination of (i) and (ii) naturally leads to the non-trivial relations between the parameters of the gravitational and matter actions.

The plan of the paper as follows. In the section \ref{section:bosonic}, we construct the geometric or gravitational part of the spectral action and compare it to the IR limit of the HL gravity. The section \ref{section:fermionic} deals with the fermionic or matter part of the spectral action. The following section discusses the possible phenomenological consequences of our approach. We collect all the notations and conventions, as well as the major part of the calculations in several appendices.

\section{Spectral action: gravity part}\label{section:bosonic}

In this section, after the brief review of the spectral action principle, we derive the gravity action based on some generalization of the standard Dirac operator. We show that this action is exactly the IR limit of the general HL gravity.

The spectral action principle \cite{Chamseddine:1996zu} deeply roots in non-commutative or spectral geometry \cite{Connes:1994yd}. The main object in this approach to geometry is the so-called spectral triple. This is strongly motivated by the reformulation of Riemannian (i.e. Euclidean) geometry of a compact manifold $M$ in terms of the commutative spectral triple: the commutative \textit{algebra} $C^\infty (M)$ acting on the \textit{Hilbert space} of square integrable spinors $L_2(S)$ and the standard \textit{Dirac operator}. In this case it was possible to prove the so-called reconstruction theorem \cite{Connes:2008vs} that proves the complete equivalence between the standard and spectral approaches. The non-commutative formulation, being purely algebraic, opens up the possibility for various generalizations of the commutative geometry. This is achieved by generalizing in some consistent way one or all of the elements of the spectral triple, see \cite{Connes:1994yd} for some examples of the generalized geometries. Physical interest in the methods of non-commutative geometry exploded after the paper \cite{Seiberg:1999vs} but even before this there had been a constant interest in physical applications. For the purposes of our work the paper \cite{Chamseddine:1996zu} is most important. In this paper the concept of the spectral action principle was introduced and later it was successfully applied to produce a new approach to the Standard Model, see e.g. \cite{Chamseddine:2008zj}.

In words, the spectral action principle can be formulated as follows: the spectral triple contains not only the complete geometrical but also physical information. The main problem is to choose the right spectral triple. If the choice is made, one \textit{postulates} that the dynamics of a physical system is governed by the following action
\begin{eqnarray}\label{action}
S = \Tr f \left( \frac{\mathbb{D}^2}{\Lambda^2} \right) + \langle J\psi , \mathbb{D}\psi \rangle \equiv S_{geom} + S_{matt} \ ,
\end{eqnarray}
where $\mathbb{D}$ is some (generalized) Dirac operator, $f$ is some cut-off function, $\Lambda$ is some characteristic scale and $J$ is a real structure (i.e. we should consider a \textit{real} spectral triple \cite{Connes:1995tu}). While the $S_{geom}$ part contains the description of gravity and gauge fields, the $S_{matt}$ term describes the natural and, in some sense, minimal interaction between the fermionic matter and the geometry (including gauge fields). As we have already mentioned, for the right choice of the real spectral triple the Standard Model perfectly fits into this scheme \cite{Chamseddine:2008zj}. A very important point about (\ref{action}) is that the same object, the generalized Dirac operator $\mathbb{D}$ controls both parts of (\ref{action}). As we briefly discussed in the introduction, this is the source of the possible relations between the parameters of two seemingly unrelated sectors. In this section we deal with the $S_{geom}$ part and discuss $S_{matt}$ in the next section.

To evaluate the trace in $\Tr f \left( \frac{\mathbb{D}^2}{\Lambda^2} \right)$ one has to use the heat kernel techniques. Here we just present the needed results, for details, see e.g. \cite{Fursaev:2011zz}. The analogous calculation for the standard Dirac operator was first done in \cite{Kastler:1993zj,Kalau:1993uc}. Because these methods (as well as the non-commutative approach to geometry) are well defined for Euclidean case, we will be working with Euclidean version of gravity. (This corresponds to the choice $\epsilon = 1$ in the calculations from the appendices.) Also we will work with a manifold without boundary.

So, let $P$ be some (generalized) elliptic operator of the order $m$ represented on a Hilbert space $\mathcal{H}$ of square-integrable sections of some bundle over d-dimensional manifold $M$. The heat kernel of $P$ is defined by
\begin{eqnarray}\label{heat_kernel}
K(t,P) = \Tr_\mathcal{H} e^{-t P} \ .\nonumber
\end{eqnarray}
This heat kernel has the well known asymptotics for small $t$
\begin{eqnarray}
K(t,P) \simeq \sum_{n\geq 0} t^{\frac{n-d}{m}} a_n (P) \ ,\nonumber
\end{eqnarray}
where $a_n (P)$ are completely defined by some local densities, $a_n (x, P)$, known as Seeley-DeWitt coefficients:
\begin{eqnarray}\label{int_a}
a_n (P) = \int_M a_n (x, P) \sqrt{g} d^d x\ .
\end{eqnarray}
There are several techniques how to calculate these coefficients one of the most effective being due to Gilkey \cite{Gilkey:1995mj}. Below we will give the explicit expressions for the first two coefficients for the case relevant for our purposes. The relation between the heat kernel and the trace of some function of $P$ can be established through the operator analog of the Mellin transformation. We write this relation for the case of our interest, i.e. when $d=4$ and $m=2$. The fact that we work with a manifold without boundary can be used to show that $a_n (P) = 0$ for all odd $n$ \cite{Fursaev:2011zz}. Then the trace of $f(P)$ is given by
\begin{eqnarray}\label{TrP}
\Tr f (P)  = \sum_{k \geq 0} f_{2k} a_{2k} (P) \ ,
\end{eqnarray}
where $f_{2k}$ are
\begin{eqnarray}
f_0  = \int\limits_0^\infty f(u) u du, \ \  f_2 =  \int\limits_0^\infty f(u) du , \ \   f_{2(k+2)} = (-1)^n f^{(n)}(0),\  k \geq 0 \ .\nonumber
\end{eqnarray}

In \cite{Chamseddine:1996zu} it was shown that choosing $P = D^2$ where $D$ is the standard Dirac operator on the twisted spinor bundle over $M$, the first three terms in (\ref{TrP}) reproduce General relativity with the cosmological constant coupled to some gauge field (depending on the twisting).

The idea of this section is to apply (\ref{TrP}) to $P = \mathbb{D}^2$, where $\mathbb{D}$ is some physically motivated deformation of $D$. As it was argued in \cite{Pinzul:2010ct,Pinzul:2014mva}, the most natural Dirac operator that respects the fundamental foliation structure of the HL geometry should be written as the most general differential operator represented on $\mathcal{H}=L_2(S)$ such that it is of the first order in time derivative and up to the third order in space derivatives. Each term should have well-defined transformation properties under FPDiff. To write such an operator explicitly, the (3+1) decomposition of the standard Dirac operator was worked out in details in \cite{Pinzul:2014mva}
\begin{eqnarray}\label{D_undeformed}
{D} = \gamma^\mu \nabla^\omega_\mu = \gamma^0 D_n + \threeD -  \frac{1}{2} \gamma^0 K + \frac{1}{2} \gamma^{\alpha} \frac{\partial_\alpha N}{N} \ .
\end{eqnarray}
Here $D_n$ is roughly speaking the covariant derivative along the normal vector, $\threeD$ is Dirac operator on the 3d leaf of the ADM-like foliation of space-time, $K$ - its extrinsic curvature and $N$ is the lapse function. (For the details of the derivation and used notations see \cite{Pinzul:2014mva} and Appendix \ref{notations} of the present paper.) While the general high order generalized Dirac operator is very hard to study,\footnote{For example, even the study of the heat kernel of the \textit{flat} case of the third order generalized Dirac operator is already technically quite involved problem \cite{Mamiya:2013wqa}. Nevertheless, in \cite{Pinzul:2010ct} it was shown that some conclusions based on the most general form of the generalized Dirac operator suitable for a HL-type theory can be made. In particular, it was shown that the UV spectral dimension will always be equal to 2.} its IR limit, i.e. when only the lowest order derivatives are kept, can be analyzed. In \cite{Pinzul:2014mva} we argued that the most general form of the deformed Dirac operator in IR limit is
\begin{eqnarray}\label{aniso-d-1}
\mathbb{D} = \gamma^0 D_n + c_1  \threeD + c_2 \gamma^0 K + c_3 \gamma^{\alpha} a_\alpha + c_4 K + c_5 \gamma^0 \gamma^{\alpha} a_\alpha \ ,
\end{eqnarray}
where $c_i, \ i=\overline{1,5}$ are some arbitrary parameters and we introduced the notation $a_\alpha = \frac{\partial_\alpha N}{N}$. Actually in \cite{Pinzul:2014mva} we used (\ref{aniso-d-1}) with $c_4=c_5 =0$. While this does not affect any of the conclusions of that paper, as we will see, adding these terms makes the real difference for the present work. (\ref{aniso-d-1}) is exactly the operator that we want to use as the fundamental object in (\ref{action}). One can bring it to a more convenient form. One of the conclusions of \cite{Pinzul:2014mva} was that from the point of view of the geodesic motion of a test particle, the original metric on $M$, $g_{\mu\nu}$, is substituted by some effective one with the re-scaled space part, $\tilde{g}_{\mu\nu}$, see Eq.(\ref{gtil}) of the appendix \ref{RRtil} with $\alpha=\frac{1}{c_1^2}$. In the appendix \ref{rescaledDirac} we show that in terms of this more natural metric (\ref{aniso-d-1}) can be re-written as (\ref{rescaled3})
\begin{eqnarray}\label{rescaled2}
\mathbb{D} = \tilde{\gamma}^\mu \nabla_\mu^{\tilde{\omega}} + \bigg( c_2 + \frac{1}{2} \bigg) K \gamma^0 + \bigg(c_3 - \frac{c_1}{2} \bigg) \gamma^i e_i^{\ \mu}a_\mu + c_4 K +  c_5 \gamma^0 \gamma^i e_i^{\ \mu}a_\mu  \ .
\end{eqnarray}
So it takes the form
\begin{eqnarray}\label{DF}
\mathbb{D} = \tilde{\gamma}^\mu \nabla_\mu^{\tilde{\omega}} + F\ ,\mathrm{\ where\ }F = \bigg( c_2 + \frac{1}{2} \bigg) K \gamma^0 + \bigg(c_3 - \frac{c_1}{2} \bigg) \gamma^i e_i^{\ \mu}a_\mu + c_4 K +  c_5 \gamma^0 \gamma^i e_i^{\ \mu}a_\mu \ .
\end{eqnarray}
To calculate the geometric part of the spectral action (\ref{action}) we need to calculate $\mathbb{D}^2$. In the appendix \ref{Lichnerowicz} we show that one can derive the generalization of the Lichnerowicz formula for the generalized Dirac operator of the form (\ref{DF})
\begin{eqnarray}\label{D2}
\mathbb{D}^2= - \tilde{g}^{\mu\nu}\nabla^{\tilde{\Omega}}_\mu \nabla^{\tilde{\Omega}}_\nu + E \ ,
\end{eqnarray}
where $\tilde{\Omega}$ and $E$ are some connection and endomorphism respectively, see eqs. (\ref{Omega}) and (\ref{E}) of the appendix \ref{Lichnerowicz} evaluated for the re-scaled metric (\ref{gtil}).
The importance of the representation (\ref{D2}) is that there is the general procedure of calculating the Seeley-DeWitt coefficients that enter in (\ref{TrP}). For our purposes we will need only the first two \cite{Gilkey:1995mj,Fursaev:2011zz}
\begin{eqnarray}\label{a_i}
a_0(x, \mathbb{D}^2) &=&  (4 \pi )^{-2} \Tr (\mathds{1}) \ ,\nonumber\\
a_2(x, \mathbb{D}^2) &=& (4 \pi )^{-2} \Tr \bigg(- \frac{\tilde{R}}{6} \mathds{1} + E \bigg)\ ,
\end{eqnarray}
where we already set that $d=4$ and the trace is over the spinor indices, i.e. in our case $\Tr (\mathds{1}) = 4$.\footnote{The dependence on $\tilde{\Omega}$ will appear only in $a_4(x, \mathbb{D}^2)$ but the contribution of this coefficient will be suppressed by $\frac{1}{\Lambda^2}$ factor.} Here $\tilde{R}$ is the 4-d curvature calculated for the re-scaled metric (\ref{gtil}). In the appendix \ref{RRtil} we show how it can be written in terms of (3+1) decomposition, see (\ref{tilR1})
\begin{eqnarray}\label{tilR2}
\tilde{R} =  c_1^2 \threeR +  K^2 -  K_{\mu\nu} K^{\mu\nu} + 2  \nabla_\mu \bigg( c_1^2 n^{\lambda} \nabla_{\lambda} n^\mu - n^\mu \nabla_{\lambda} n^\lambda \bigg) \ .
\end{eqnarray}
The trace of the endomorphism $E$ is calculated in the appendix \ref{Lichnerowicz} (\ref{TrE}). To apply this result to our Dirac operator (\ref{DF}) we have to identify the coefficients $a$, $b_\mu$ and $c_{\mu\nu}$ from the appendix formula (\ref{F}) with the parameters in (\ref{DF}):
\begin{eqnarray}
a =  c_4 K  \ ,\ \ b_a = \left( \bigg( c_2 + \frac{1}{2} \bigg) K\ ,\ \bigg(c_3 - \frac{c_1}{2} \bigg) e_i^{\ \mu}a_\mu\right)\ ,\ c_{0i} = \frac{1}{2}c_5 e_i^{\ \mu}a_\mu \ ,\ c_{ij} = 0 \ ,\nonumber
\end{eqnarray}
where we passed from the space-time indices to the flat (tangent) ones. Then (\ref{TrE}) will become
\begin{eqnarray}\label{TrE1}
\Tr E &=&  -12 c_4^2 K^2 + 4 c_5^2 a^\mu a_\mu + \tilde{R}\ .
\end{eqnarray}
Using (\ref{tilR2}) and (\ref{TrE1}) in (\ref{a_i}) we can easily get the integrated Seeley-DeWitt coefficients (\ref{int_a})
\begin{eqnarray}\label{int_a_i}
a_0(\mathbb{D}^2) &=&  \frac{1}{4\pi^2} \int_M  \sqrt{g} d^4 x \ ,\nonumber\\
a_2(\mathbb{D}^2) &=& \frac{1}{48\pi^2}  \int_M \left(  c_1^2\threeR + (1-36 c_4^2) K^2 - K_{\mu\nu} K^{\mu\nu} + 12 c_5^2 a^\mu a _\mu \right) \sqrt{g} d^4 x \ .
\end{eqnarray}
Finally, using (\ref{int_a_i}) in (\ref{TrP}) we arrive at the expression for the geometric part of the spectral action $S_{geom}$
\begin{eqnarray}
S_{geom} = \frac{f_2 \Lambda^2}{48\pi^2}  \int_M \left( c_1^2 \threeR + (1-36 c_4^2) K^2 - K_{\mu\nu} K^{\mu\nu} + 12 c_5^2 a^\mu a _\mu + \frac{12 f_0 \Lambda^2}{f_2} \right) \sqrt{g} d^4 x \ .\nonumber
\end{eqnarray}
This should be compared with (the Euclidean version of) the IR limit of the action for HL gravity with the cosmological constant $\Lambda_c$ \cite{Barausse:2011pu,Barausse:2013nwa}
\begin{eqnarray}\label{IRHL}
S_{IRHL} = \frac{M_P^2}{2} \int_M \left( \xi \threeR + \lambda K^2 - K_{\mu\nu} K^{\mu\nu} + \eta a^\mu a _\mu + \Lambda_c \right) \sqrt{g} d^4 x \ .
\end{eqnarray}
It is obvious that upon the following identifications of the free parameters:
\begin{eqnarray}\label{eq:compare healthy}
M_P^2 =\frac{f_2 \Lambda^2}{24\pi^2}\ ,\ \  \sqrt{c_1} = \xi\ ,\ \  \lambda = (1-36 c_4^2)\ ,\ \ \eta = 12 c_5^2\ ,\ \ \Lambda_c = \frac{12 f_0 \Lambda^2}{f_2}
\end{eqnarray}
both actions are exactly the same. This is the main result of this section. Using the map between the parameters (\ref{eq:compare healthy}) we can translate the experimental bounds on $(\xi , \lambda , \eta)$ into the bounds on $(c_1 , c_4 , c_5)$, see below.  It is important to note that without the P-violating terms in (\ref{aniso-d-1}), which are proportional to $c_4$ and $c_5$, the resulting action would be a rather trivial deformation of GR. Also note that up to this order there are no bounds on $(c_2 , c_3)$. This should be compared to the same conclusion in \cite{Pinzul:2014mva}. This situation changes when we include matter. So we pass to the consideration of the matter part of the spectral action (\ref{action}), $S_{matt}$.

\section{Spectral action: matter part}\label{section:fermionic}

The spectral action principle (\ref{action}) tells us that the action for fermionic matter is given by
\begin{equation}\label{Smatter}
S_{matter} = \langle J\psi , \mathbb{D}\psi \rangle \equiv \int d^4 x \sqrt{g} \bar \psi \mathbb{D} \psi \ .
\end{equation}

Before we proceed with our analysis of (\ref{Smatter}) some comments are in order. The matter in HL gravity has been considered before with the conclusion that to avoid problems (fine tuning, strong Lorentz violation, etc) one has to consider the minimal coupling of matter sector to the geometry, i.e. the same as in GR \cite{Pospelov:2010mp,Kimpton:2013zb}. We can immediately see that this is against the spirit of the spectral action principle - clearly the matter in (\ref{Smatter}) is not minimally coupled in this sense. But it \textit{is} minimally coupled from the point of view of the spectral action - the action (\ref{Smatter}) is really minimal and the most natural action built of some physical Dirac operator. The fine tuning, as we said before, could be if not avoided completely but at least improved by the fact that the same Dirac operator controls both parts of the full action (see also the discussion below).

Recalling the form of our generalized Dirac operator (\ref{aniso-d-1}) it should be obvious that the action (\ref{Smatter}) will lead to Lorentz violation, that can in principle be detectable in experiments, which, in its turn, will put severe constrains on the parameters in (\ref{aniso-d-1}). In \cite{Kostelecky:2003fs}, a general model independent Lorentz violating extension of Standard Model (SME) was presented and in \cite{Kostelecky:2008ts}, the coefficients appearing in \cite{Kostelecky:2003fs} are compared to precision experiments. Our goal here is to see how (\ref{Smatter}) fits into this scheme. As we have stressed many times the same free parameters $c_i$ enter both the geometric (gravitational) and fermionic (matter) parts of the action. So if we take the spectral action principle seriously, then we can use the data from cosmology or gravitational precision measurements to constrain Standard Model deviations from Lorentz invariance, and vice-versa. Of course to do so one has to work with the complete SM (with all the gauge field content). We will use our (somewhat toy) model to demonstrate how this should work in principle.

To proceed, let us write down the most general SME action in the fermionic sector \cite{Kostelecky:2003fs}
\begin{eqnarray}\label{SME}
S=\int d^4 x \sqrt{g} ( e^{\ \mu}_a \bar \psi \Gamma^a \nabla_\mu \psi + \bar \psi M \psi) \ ,
\end{eqnarray}
where
\begin{eqnarray}\label{LVT}
\Gamma^a &=&\gamma^a -  c_{\mu\nu} e^{ a\nu} e^{\ \mu}_b \gamma^b -  d_{\mu\nu} e^{ a\nu}e^{\ \mu}_b \gamma^5 \gamma^b - e_\mu e^{ a\mu} - i f_\mu e^{ a\mu} \gamma^5 - \frac{1}{2} g_{\lambda\mu\nu}e^{ a\nu}e^{\ \lambda}_b e^{\ \mu}_c \gamma^{bc} \ ,\nn \\
M &=& m + i m_5 \gamma^5 + m_\mu e_a^{\ \mu} \gamma^a +  b_\mu e^{\ \mu}_a \gamma^5 \gamma^a + \frac{1}{2}  H_{\mu\nu} e^{\ \mu}_a e^{\ \nu}_b \gamma^{a b} \ .
\end{eqnarray}
Because (\ref{SME}) is model independent, all the parameters in (\ref{LVT}) are independent. If one tries to arrive at (\ref{SME}) within a specific model, typically there will be some conditions on these parameters. E.g. this happens when one re-writes the noncommutative electrodynamics in SME form, see \cite{Chatillon:2006rn}. To see how this works in our case we have to re-write (\ref{Smatter}) in the form (\ref{SME}). Here, actually, we have two alternatives.

i) Using (\ref{rescaled1}), we can easily show that the Dirac operator (\ref{aniso-d-1}) can be written in the form
\begin{eqnarray}\label{Def1}
\mathbb{D} = \gamma^\mu \nabla^\omega_\mu + (c_1 -1) e_i^{\ \mu} \gamma^i \nabla^\omega_\mu + \left(\frac{c_1}{2} + c_2 \right) \gamma^0 K + \left(c_3 - \frac{1}{2}\right) \gamma^i e_i^{\ \mu}a_\mu + c_4 K +  c_5 \gamma^0 \gamma^i e_i^{\ \mu}a_\mu \ .
\end{eqnarray}

ii) Alternatively, we can directly use the representation (\ref{rescaled2})
\begin{eqnarray}\label{Def2}
\mathbb{D} = \tilde{\gamma}^\mu \nabla_\mu^{\tilde{\omega}} + \bigg( c_2 + \frac{1}{2} \bigg) K \gamma^0 + \bigg(c_3 - \frac{c_1}{2} \bigg) \gamma^i e_i^{\ \mu}a_\mu + c_4 K +  c_5 \gamma^0 \gamma^i e_i^{\ \mu}a_\mu  \ .
\end{eqnarray}

Naively, it would seem that (\ref{Def1}) is the right choice because its first term represents the standard Dirac operator with respect to the original metric $g_{\mu\nu}$. But taking into account the conclusion of the paper \cite{Pinzul:2014mva} that the physical geodesic motion of a test particle is defined as the usual geodesic motion but with respect to the re-scaled $\tilde{g}_{\mu\nu}$ metric, the choice (\ref{Def2}) might seem more natural physically. In any case, this requires some further understanding and here we treat both cases.

i) Comparing the first choice (\ref{Def1}) to (\ref{SME},\ref{LVT}) we easily identify the non-vanishing coefficients
\begin{eqnarray}\label{set1}
c_{\mu \nu} = (c_1 -1) h_{\mu \nu}  \ ,\  m_\mu = \left(\left(\frac{c_1}{2} + c_2 \right) K \ ,\ \left(c_3 - \frac{1}{2}\right) a_\alpha \right)  \ ,\ m = c_4 K \ ,\ H_{0 \alpha} = 2 c_5 a_\alpha \ .
\end{eqnarray}
To get $c_{\mu \nu}$ we used (\ref{projector1}).

ii) Doing the same for the choice (\ref{Def2}) we immediately arrive at the following set of the Lorentz violating coefficients:
\begin{eqnarray}\label{set2}
m_\mu = \left(\left(\frac{1}{2} + c_2 \right) K \ ,\ \left(c_3 - \frac{1}{2}\right) a_\alpha\right)  \ ,\ m = c_4 K \ ,\ H_{0 \alpha} = 2 c_5 a_\alpha \ ,
\end{eqnarray}
the rest of the coefficients being zero.

So the difference between two cases is minimal and essentially reduces to the absence of the $c_{\mu \nu}$ correction in the second case, which is predictable due to the metric re-scaling.

There are several short comments that we would like to make before proceeding with the discussion of the obtained results.

Firstly, the matter part of the spectral action finally gives some non-trivial dependence on the coefficients $c_2$ and $c_3$. As we have seen in the section \ref{section:bosonic} these coefficients completely drop out of the geometric action. This, of course, agrees with the earlier observation that the corrections to the physical geodesic motion due to the generalized nature of (\ref{aniso-d-1}) depend only on $c_1$.

Secondly, it is important to notice that all the Lorentz violating coefficients depend on the background geometry. This is very important when comparing to experiments because the absence of the measurable corrections would not automatically mean that they are small in general but could be just because the gravity is weak in this particular experiment (see also below). As an example of such correction we can give the gravity dependent mass: even if the fermionic field has a vanishing bare mass $m=0$, the term
\begin{eqnarray}
S_{\text{mass}} = \int d^4 x \sqrt{g} \bar \psi c_4 K \psi \nonumber
\end{eqnarray}
will generate an effective mass different from zero, dependent on the extrinsic curvature of the space.

Thirdly, from (\ref{set1}) or (\ref{set2}) it is clear that, of course, the natural way to parameterize the Lorentz violating sector is not by $c_i$ but by the deviations of these coefficients from the undeformed values $\{c_i \}= (1, -1/2, 1/2, 0, 0)$ (this can be easily derived by comparing (\ref{D_undeformed}) and (\ref{aniso-d-1})).

\section{Constraining the parameters}

Typically the bounds on the parameters of Horava-Lifshitz gravity are coming from the experimental tests where gravitational effects are significant. Usually, the experiments in principle capable of testing the theory do not have enough precision do differentiate from usual GR or are still unrealizable (see for example \cite{Lobo:2010hv,Barausse:2011pu,Barausse:2013nwa} and references therein). If however the spectral action formalism is the correct one to deal with HL gravity, than the parameters of the gravitational action are related to the parameters in the fermionic action and those are constrained by  much more feasible experiments. \\

As described in the previous section, in the  spectral action formalism the parameters of the effective IR theory for the bosonic and fermionic degrees of freedom are both obtained from the same object, the anisotropic Dirac operator. When we choose the Dirac operator \eqref{aniso-d-1} we reproduce, to the lowest order, the IR limit of Horava-Lifshitz theory and a Lorentz violating fermionic action. In \cite{Kostelecky:2003fs}, the general Lorentz violating extension of QED was presented \eqref{SME}, and in \cite{Kostelecky:2008ts} the experimental bounds for the general parameters in (22) were discussed. These parameters are related to the parameters we introduced for the Lorentz violating Dirac operator \eqref{Def1} or \eqref{Def2}
via \eqref{set1} or \eqref{set2} respectively. The same parameters reproduce the Horava-Lifshitz action (to the leading order) via \eqref{eq:compare healthy}. Thus, combining \eqref{eq:compare healthy} and \eqref{set1}, we get\footnote{If we instead use the Dirac operator with the rescaled metric \eqref{Def2}, we will get for (\ref{relationmattergravityb}) below
	\[
	c_{\mu\nu} = 0 , \  m_\mu = \left(\left(\frac{1}{2}+c_2 \right) K\ , \ \left(c_3 -\frac{1}{2}\right) a_\alpha\right)\ , \ m=\frac{1}{6}\sqrt{\frac{\beta-\mu}{\beta-1}}K\ , \ H_{0\alpha} = \sqrt{\frac{\alpha}{3-3\beta}} a_\alpha \ . \
	\] To avoid cluttering in the following analysis we consider only the first case, but all the arguments follow analogously.}
\begin{equation}\label{relationmattergravity}
c_{\mu \nu} = (\xi^2 -1) h_{\mu \nu}\ , \  m_\mu = \left(\left(\frac{\xi^2}{2}+c_2 \right) K\ , \ \left(c_3 -\frac{1}{2}\right) a_\alpha\right)\ , \ m=\frac{1}{6}\sqrt{1-\lambda}\,K\ , \ H_{0\alpha} = \sqrt{\frac{\eta}{3}} a_\alpha \ .
\end{equation}
Therefore, the experimental bounds on the particle side relate directly to the gravity side.

We can rewrite \eqref{relationmattergravity} using parameters $(\beta, \mu, \alpha)$ more convenient to compare with astrophysical data \cite{Barausse:2013nwa}
\begin{equation}\label{HLparameters}
\xi = \frac{1}{1-\beta}\ , \ \lambda = \frac{1-\mu}{1-\beta}\ , \ \eta = \frac{\alpha}{1-\beta}\ ,\nonumber
\end{equation}
so ($m_\alpha$ is unchanged)
\begin{equation} \label{relationmattergravityb}
c_{\mu \nu} = \left[\frac{1}{(\beta-1)^2}-1\right]  h_{\mu \nu}\ , \ m_0 = \left[\frac{1}{2(\beta-1)^2}+c_2 \right] K\ , \
m=\frac{1}{6}\sqrt{\frac{\mu - \beta}{1-\beta}}K\ , \ H_{0\alpha} = \sqrt{\frac{\alpha}{3-3\beta}} a_\alpha \ .
\end{equation}

It is instructive to write the relations above as a table connecting the Lorentz violating parameters as in \cite{Kostelecky:2003fs}, the parameters of the anisotropic Dirac operator \eqref{Def1} $c_i$ or the IR Horava-Lifshitz coefficients $(\xi, \lambda, \eta)$ and the ADM functions (lapse function, 3d metric and extrinsic curvature):

\begin{table}[h!]
	\centering
\begin{tabular}{c | c  | c}
LV parameter & HL/anis-Dirac parameters & ADM functions \\
	$c_{\mu \nu}$ & $\xi$ (or $\beta$)& $h_{\mu \nu}$ \\
	$m_0$ & $\xi$ (or $\beta$), $c_2$ & $K$ \\
	$m_\alpha$ & $c_3$ & $a_\alpha$ \\
	$m$   & $\lambda$ (or $\beta$ and $\mu$) & $K$ \\
	$H_{0\alpha}$ & $\eta$ (or $\beta$ and $\alpha$)& $a_\alpha$
\end{tabular}
\end{table}

The spectral action formalism provides us a connection among the three columns. The first column exhibits Lorentz violating parameters for the action describing matter (fermionic) fields. Experiments testing the propagation and dynamics of fermions can provide bounds on these parameters, even without gravitational effects. In \cite{Kostelecky:2008ts} it is presented many such experiments and bounds.

The second column contais the parameters that measure deviations of HL gravity from standard GR, and are therefore constrained by gravitational experiments, i.e., experiments where the gravitational field is significant. For example, in \cite{Barausse:2011pu,Barausse:2013nwa} it is presented how measurements of the radiation emitted by the accretion disks around black-holes or gravitational lensing can provide bounds on the HL parameters. In \cite{Lobo:2010hv} Solar System tests (radar echo, perihelion precession of Mercury and light deviation by the Sun) are presented.

The third column is related to the gravitational field and the foliation structure of spacetime and thus, in principle, can be obtained directly from theoretical calculations. For example, in appendix \ref{ap:spherical} we show how to write $a_\alpha$ in the next-to-leading order in Horava-Lifshitz parameters for the case of a weak field for earth-based experiments and obtain the explicit form for the radial component of the Lorentz violating parameter $m_\mu$ (\ref{m_r})
\begin{equation}
m_r = \left( c_3 - \frac{1}{2}\right) \frac{G_N M}{ r^2} \left( 1 + \frac{G_N^2 M^2}{4 r^2}  \frac{\eta}{\xi} + \ldots \right) .\nonumber
\end{equation}

As an example of how to connect the three columns presented above, consider the experimental constrain on the HL parameter $\beta$. In  \cite{Barausse:2011pu,Barausse:2013nwa}, it is argued that by analyzing the frequencies and redshifts of X-ray emitted by accretion disks of black-hole candidates,\footnote{GR modifications change the Innermost Stable Circular Orbit (ISCO), and therefore modify accretion disks.} their emission lines and the impact parameter of the circular photon orbit (that affects gravitational lensing) one can in principle measure possible Horava-Lifshitz deviations from the standard Einstein gravity. However, these deviations from GR are too small to be conclusively measured (at least at the present time). For instance, a large value of $\beta=0.2$ (that is, $\xi$=1.25) gives a deviation of the redshift measured in accretion disks by less than $0.1 \% $ when compared to the GR prediction.

If however we take the spectral action formalism as the correct way to deal with effective actions, we can instead look at experiments in the matter sector where we have much better precision. Altschul \cite{Altschul:2010na} gives bounds on the electron Lorentz violating parameters $c_{IJ}$ by analyzing synchrotron radiation and inverse Compton scattering from astrophysical sources, with $I,J=X,Y,Z$. Here the coordinate basis is chosen in a way that $Z$ corresponds to the Earth north, i.e. parallel to Earth rotation axis, $X$ corresponds to the direction from the Earth to the Sun at the vernal equinox, and $Y$ follows the right-hand rule. E.g., the bound on $c_{XX}$ is given by

\begin{equation}\label{betagrav}
-5 \times 10^{-15} \lesssim c_{XX} \lesssim 5 \times 10^{-15} .
\end{equation}

In the spectral action formalism $c_{XX}$ is related to $\beta$ via \eqref{relationmattergravityb}
\begin{equation}
c_{XX} = \left[\frac{1}{(\beta-1)^2}-1\right]  h_{XX} \ . \nonumber
\end{equation}

These electrons are produced in different astrophysical sources, so for each one we would have to take into account $h_{XX}$ in order to calculate $\beta$. If we consider sources where the gravitational field is weak, approximating $h_{XX} \sim 1$ we get

\begin{equation}\label{betapart}
|\beta| \lesssim 2.5 \times 10^{-15}\ ,
\end{equation}
which is a much stronger bound then the one coming from any of the gravitational experiments. Thus, while the still challenging gravitational experiments may not provide us with bounds on GR modifications, other kinds of high precision experiments may.

\section{Discussion and conclusion}

In this work we studied the application of the spectral action principle (\ref{action}) to the gravity plus matter theory based on the foliation preserving diffeomorphisms. Choosing (\ref{aniso-d-1}) as the physically motivated generalized Dirac operator in the infrared regime, we were able to reproduce the most general IR action for HL gravity coupled in a non-trivial way to fermionic matter.

One of the main conclusions that can be made based on our results is that, if taken seriously, the spectral action principle can serve as an effective tool in reducing the parameter space of the theory. E.g., in the IR regime, considered in this work, the parameter space is just 5-dimensional one, spanned by $\{c_i\}$.

Another very important (and potentially could be the most important) advantage of our approach, intimately related to the previous one, is the possibility to use some high precision experiments from one sector (typically this would be the matter sector) to constrain the parameters of the other one. We demonstrated how this could be done in principle on the example of the parameter $\beta$, which does not have very restrictive bounds coming from gravitational experiments, see (\ref{betagrav},\ref{betapart}).

We have not done any serious confrontation of our model with the existing experimental data. This is because our model is still a toy model that served well to our purpose - demonstration of the advantages and the potential of our approach. Of course, the most obvious and urging next step is to make our model less a toy, which will allow to look at the experimental bounds on parameters of the model. This will require adding the gauge sector to our construction. The spectral action principle provides a natural way for the inclusion of gauge fields - they will become a part of a generalized Dirac operator on some appropriately twisted spinor bundle. The example for the usual case is considered at the end of the appendix \ref{Lichnerowicz}. For the case of a theory based on FPDiffs, one has to construct the twisting consistent with this symmetry. This problem is currently under consideration.

As we stressed several times, our approach could help to avoid some fine tunings needed, as it was argued in \cite{Kimpton:2013zb}, for the consistent non-minimal coupling of matter. To check whether this is the case, one has to do the analysis along the lines of \cite{Kimpton:2013zb} but now with the constrained parameter space. The optimistic expectation would be that some problems observed in \cite{Kimpton:2013zb} will be cured by the built in relations between the parameters. We are planning to address this problem in the nearest future.

Much more ambitious and fundamental problem is the generalization of our approach to the full theory, i.e. not just to the IR limit. As we commented above this would require working with the most general deformation of Dirac operator by the terms up to the third order in space derivatives. This is a very challenging, both technically and conceptually, problem. Some initial steps were taken in \cite{Mamiya:2013wqa}, but a lot of work still should be done before one could try to derive the full HL theory from the spectral action principle. Taking into account the advantages of this approach, we still hope that this will be possible to accomplish.

\section*{Acknowledgements}

A.P. acknowledges partial support of CNPq under grant no.306068/2012-5. The major part of the work was done while D.V.L. was supported by the CNPq grant no.501107/2012-6 and A.M. by CNPq master’s scholarship.

\begin{appendices}

\section{Notations and conventions}\label{notations}

Here we fix the notations and conventions used in the main text as well as establish some formulas used in the further calculations. Because this is very standard material, we essentially copy one of the appendices of \cite{Pinzul:2014mva}. Though we will need only the Euclidean signature, for the sake of generality we will obtain our results for both, Euclidean ($\epsilon = 1$) and Lorentzian ($\epsilon = -1$), cases. So the only difference from \cite{Pinzul:2014mva} is that now we explicitly write all formulas for both signatures and use the algebra of $\gamma$-matrices that differs by a sign from \cite{Pinzul:2014mva}.

\textit{Coordinate system.} We adopt the coordinates compatible with the foliation structure:
\begin{eqnarray}\label{coord}
x^\mu = (t, \vec{x})\ ,
\end{eqnarray}
where $t=const$ defines a leaf of the foliation $\Sigma_t$, while $\vec{x}$ are the coordinates on $\Sigma_t$.

\textit{Metric.} In the coordinates (\ref{coord}), the metric takes the usual ADM form \cite{Arnowitt:1962hi}
\begin{eqnarray}\label{ADM}
d s^2 = \epsilon (Ndt)^2 + h_{\alpha\beta} (dx^\alpha + N^\alpha dt)(dx^\beta + N^\beta dt)\ ,
\end{eqnarray}
where $N$ is the laps function and $N^\alpha$ is the shift vector and $h_{\alpha\beta}$ is the 3d metric on a leaf. Throughout the paper we are using the following system of indices:

The Greek letters from the middle of the alphabet, $\mu,\ \nu,\ \ldots $ are used to denote the curved coordinates (\ref{coord}) and take values 0, 1, 2 and 3.

The Greek letters from the beginning of the alphabet, $\alpha,\ \beta,\ \ldots $ are used to denote the space part of the curved coordinates (\ref{coord}), i.e. $\vec{x}$, and take values 1, 2 and 3.

The Latin letters from the beginning of the alphabet, $a,\ b,\ \ldots $ are used to denote the coordinates of 4d flat space and take values 0, 1, 2 and 3.

The Latin letters from the middle of the alphabet, $i,\ j,\ \ldots $ are used to denote the coordinates of the space part of 4d flat space and take values 1, 2 and 3.

\textit{Tetrads, Second fundamental form.} We partially fix the local $SO(3,1)$ or $SO(4)$ invariance (which is natural to do keeping in mind the fundamental meaning of the foliation) and choose $e_0^{\ \mu}$ to be equal to the vector normal to $\Sigma_t$, i.e.
\begin{eqnarray}\label{zerotetrad}
e_0^{\ \mu} = n^\mu\ ,
\end{eqnarray}
where $n^\mu$ is the vector dual to the 1-form $n=\epsilon N dt$. Clearly, this vector is normal to the hypersurface $t=const$, and using (\ref{ADM}) we see that
\begin{eqnarray}\label{normal}
n_\mu=(\epsilon N,0,0,0)\ , \ n^\mu = \left(\frac{1}{N},\epsilon\frac{N^\alpha}{N}\right)\ \mathrm{and} \ n^\mu n_\mu = \epsilon \ .\nonumber
\end{eqnarray}
Then the rest of the tetrads will belong to the space tangent to $\Sigma_t$, $e_i^{\ \mu}\in \mathrm{T}\Sigma_t$. As usual, we can introduce the projector on $\mathrm{T}\Sigma_t$
\begin{eqnarray}\label{projector}
h_{\mu\nu} = g_{\mu\nu} - \epsilon n_\mu n_\nu \ .
\end{eqnarray}
The fact that $h_{\mu\nu}$ projects any vector from $\mathrm{T}M$ to a vector in $\mathrm{T}\Sigma_t$, immediately follows from that a) $h_{\mu\nu} n^\mu = 0$ and b) $h_{\mu}^{\ \rho} h_{\rho\nu} = h_{\mu\nu}$. Using this, one can easily see that $e_i^{\ \mu}$ are left invariant by $h_{\mu\nu}$, $h^{\mu}_{\ \nu} e_i^{\ \nu} = e_i^{\ \mu}$. Also, combining (\ref{zerotetrad}), (\ref{projector}) and $e_{a\mu} e^{a}_{\ \nu}=g_{\mu\nu}$, one can establish that
\begin{eqnarray}\label{projector1}
h_{\mu\nu} = e_{i\mu} e^{i}_{\ \nu} \ .
\end{eqnarray}

Using the normal vector $n^\mu$ and the projector $h_{\mu\nu}$ we define in the standard way a second fundamental form, or the extrinsic curvature, which measures how the leaves of the foliaton are ``bended'' in the ambient space-time
\begin{eqnarray}\label{externalcurvature}
K_{\mu\nu} = -h_\mu^{\ \rho}\nabla_\rho n_\nu\ \ .
\end{eqnarray}
From this definition and using (\ref{projector}) we can establish a very useful relation:
\begin{eqnarray}\label{nabla_n}
\nabla_\mu n_\nu = \epsilon n_\mu n^\sigma \nabla_\sigma n_\nu - K_{\mu\nu}\ .\nonumber
\end{eqnarray}

\textit{Covariant spin derivative.} The covariant derivative on a spin bundle is defined in the usual way, e.g. \cite{Nakahara:1990th}:
\begin{eqnarray}\label{spincovder}
\nabla^\omega_\mu = \partial_\mu + \omega_\mu \ ,
\end{eqnarray}
where $\omega_\mu = -\frac{1}{4}\omega_{\mu ab}\gamma^{ab}$ is a spin connection and $\gamma^{ab}:= \frac{1}{2}[\gamma^{a},\gamma^{b}]$ are the generators of $SO(3,1)$ or $SO(4)$. Here $\gamma^{a}$ are the usual flat gamma matrices, i.e. $\{\gamma^{a},\gamma^{b}\} = -2\eta^{ab}$, where $\eta^{ab} = \mathrm{diag}(\epsilon,1,1,1)$. The condition that covariant derivative is compatible with metric is translated into the full (i.e. with respect to both space-time and flat indices) covariant constancy of the tetrad:
\begin{eqnarray}
\tilde{\nabla}_\mu e_{a\nu} \equiv \partial_\mu e_{a\nu} + \omega_{\mu a}^{\ \ b}e_{b\nu} - \Gamma^{\rho}_{\mu\nu}e_{a\rho} = 0\ .
\end{eqnarray}
From here, it is easy to find the expression for $\omega_{\mu ab}$
\begin{eqnarray}\label{spinconnection}
\omega_{\mu ab} = e_{a\nu}\partial_\mu e_{b}^{\ \nu} + \Gamma^{\rho}_{\mu\nu}e_{a\rho}e_{b}^{\ \nu}\equiv e_{a\nu}\nabla_\mu e_{b}^{\ \nu}\ ,
\end{eqnarray}
where now $\nabla_\mu$ is the usual space-time covariant derivative, i.e. the one acting on space-time indices only.

\textit{Dirac operator.} We define the standard Dirac operator without $\sqrt{-1}$ in front:
\begin{eqnarray}\label{Dirac}
\mathrm{D} = \gamma^\mu \nabla^\omega_\mu \ ,
\end{eqnarray}
where $\gamma^\mu := e_{a}^{\ \mu}\gamma^a$ are the curved gamma matrices, i.e. $\{\gamma^\mu,\gamma^\nu\} = -2g^{\mu\nu}$ (which is trivial by $e_{a\mu} e^{a}_{\ \nu}=g_{\mu\nu}$).

\section{Generalized Lichnerowicz formula}\label{Lichnerowicz}

Here we would like to calculate ${\mathbb{D}}^2$ for a generalized Dirac operator ${\mathbb{D}} = \gamma^\mu \nabla_\mu + F$, where $\nabla_\mu = \partial_\mu + \omega_\mu$ is a metric compatible covariant derivative on a spinor bundle and $F$ is some arbitrary endomorphism of this bundle. Later on we will specify the form of $F$. Essentially the goal is achieved in two steps: 1) Bring ${\mathbb{D}}^2$ to the form ${\mathbb{D}}^2 = -g^{\mu\nu}\partial_\mu \partial_\nu + A^\mu \partial_\mu + B$, where $A$ and $B$ are some endomorphisms of the bundle;
2) Using the well-known fact, see e.g. \cite{Fursaev:2011zz}, that in this case there always exists a connection $\Omega_\mu$ and an endomorphism $E$ such that $-g^{\mu\nu}\partial_\mu \partial_\nu + A^\mu \partial_\mu + B = -g^{\mu\nu}\nabla^\Omega_\mu \nabla^\Omega_\nu + E$, find these $\Omega_\mu$ and $E$.

1) ${\mathbb{D}}^2 = - g^{\mu\nu}\partial_\mu \partial_\nu + A^\mu \partial_\mu + B$
\begin{eqnarray}
{\mathbb{D}}^2 &=& (\gamma^\mu \nabla_\mu + F)(\gamma^\nu \nabla_\nu + F) = \gamma^\mu \nabla_\mu \gamma^\nu \nabla_\nu +\gamma^\mu [\nabla_\mu , F] +\{\gamma^\mu , F \} \nabla_\mu + F^2 = \nonumber \\
&=& -{g}^{\mu\nu}\partial_\mu \partial_\nu - (2{g}^{\mu\nu} \omega_\nu - \{\gamma^\mu, F \} - \Gamma^\mu )\partial_\mu -\nonumber \\
&& - {g}^{\mu\nu} (\partial_\nu \omega_\mu  +  \omega_\mu \omega_\nu - {\Gamma}^\sigma_{\mu\nu} \omega_\sigma ) + \frac{{R}}{4} + {\gamma^{\mu}} ( \partial_{\mu} F) +  \{ {\gamma^{\mu}} \omega_\mu , F \} + F^2 \ ,\nonumber
\end{eqnarray}
where $\Gamma^\mu = {g}^{\sigma\nu}{\Gamma}^\mu_{\sigma\nu}$ and we used the usual Lichnerowicz formula: $\gamma^\mu \nabla_\mu \gamma^\nu \nabla_\nu = -\nabla^\mu \nabla_\mu +\frac{{R}}{4}$. From this result we identify $A^\mu$ and $B$:
\begin{eqnarray}\label{AB}
A^\mu &=& - 2{g}^{\mu\nu} \omega_\nu + \{\gamma^\mu, F \} + \Gamma^\mu \nonumber \\
B &=& - {g}^{\mu\nu} (\partial_\nu \omega_\mu  +  \omega_\mu \omega_\nu - {\Gamma}^\sigma_{\mu\nu} \omega_\sigma ) + \frac{{R}}{4} + {\gamma^{\mu}} ( \partial_{\mu} F) +  \{ {\gamma^{\mu}} \omega_\mu , F \} + F^2 \ .\nonumber
\end{eqnarray}

2) ${\mathbb{D}}^2 = - g^{\mu\nu}\nabla^\Omega_\mu \nabla^\Omega_\nu + E$

This step is standard (see e.g. \cite{Fursaev:2011zz,Kalau:1993uc}). Defining
\begin{eqnarray}
\Omega_\mu &=& - \frac{1}{2} g_{\mu\nu}(A^\nu - \Gamma^\nu) \ , \nonumber \\
 E &=& B + {g}^{\mu\nu} (\partial_\nu \Omega_\mu  +  \Omega_\mu \Omega_\nu - {\Gamma}^\sigma_{\mu\nu} \Omega_\sigma )\nonumber
\end{eqnarray}
one easily establishes the desired relation: $-g^{\mu\nu}\partial_\mu \partial_\nu + A^\mu \partial_\mu + B = -g^{\mu\nu}\nabla^\Omega_\mu \nabla^\Omega_\nu + E$. In our case we have from (\ref{AB})
\begin{eqnarray}\label{Omega}
\Omega_\mu &=& \omega_\mu - \frac{1}{2} {g}_{\mu\nu} \{{\gamma}^\nu , F \} =: \omega_\mu - I_\mu\ , \nonumber \\
 E &=& {\gamma^{\mu}} ( \partial_{\mu} F)  +  \{ {\gamma^{\mu}} \omega_\mu , F \} + F^2 - \nonumber \\
 &&  - {g}^{\mu\nu}  \partial_\mu I_\nu - {g}^{\mu\nu} \{ \omega_\mu, I_\nu \} + {g}^{\mu\nu}  I_\mu I_\nu + \tensor{{\Gamma}}{^\mu} I_\mu + \frac{{R}}{4} \ .
\end{eqnarray}
Using that $\nabla_\mu F = \partial_\mu F + [\omega_\mu , F]$ and $\nabla_\mu I_\nu = \partial_\mu I_\nu + [\omega_\mu , I_\nu] - \Gamma^\sigma_{\mu\nu} I_\sigma$, the expression for $E$ is simplified to become
\begin{eqnarray}\label{E}
E &=& \frac{1}{2} [ {\gamma}^\mu, \nabla_\mu F] + F^2 + \frac{1}{4}{g}_{\mu\nu} \{{\gamma}^\mu , F \}\{{\gamma}^\nu , F \} + \frac{{R}}{4} \ .
\end{eqnarray}

For our purposes, we need to calculate the trace of $E$. Clearly the trace of the commutator in (\ref{E}) is zero. So the only non-trivial trace is $\Tr(F^2 + \frac{1}{4}{g}_{\mu\nu} \{{\gamma}^\mu , F \}\{{\gamma}^\nu , F \})$. Using the cyclic property of the trace and the algebra of the 4d gamma matrices, this reduces to $-\Tr(F^2 - \frac{1}{2}{g}_{\mu\nu} {\gamma}^\mu F {\gamma}^\nu  F )$. To evaluate this trace, we will take the most general form of $F$
\begin{eqnarray}\label{F}
F = a + \gamma^\mu b_\mu + \gamma^{\mu\nu} c_{\mu\nu}	\ ,
\end{eqnarray}
where $a,\ b_\mu$ and $c_{\mu\nu}$ are some functions. Because the trace of a product of an odd number of gamma matrices is always zero and using the well-known trace
\begin{eqnarray}
\Tr( \gamma^\mu \gamma^\nu \gamma^\rho \gamma^\sigma ) = 4 (g^{\mu\nu}g^{\rho\sigma}-g^{\mu\rho}g^{\nu\sigma} + g^{\mu\sigma} g^{\nu\rho})\nonumber
\end{eqnarray}
we get
\begin{eqnarray}
\Tr F^2 &=& \Tr(a^2 - b^\mu b_\mu + c_{\mu\nu}c_{\rho\sigma}\gamma^{\mu\nu}\gamma^{\rho\sigma})= 4(a^2 - b^\mu b_\mu - 2 c_{\mu\nu}c^{\mu\nu}) \nonumber\\
{g}_{\mu\nu} \Tr ( {\gamma}^\mu F {\gamma}^\nu  F ) &=& {g}_{\mu\nu} \Tr ({\gamma}^\mu {\gamma}^\nu a^2 - \gamma^\mu \gamma^\rho \gamma^\nu \gamma^\delta b_\rho b_\delta + \gamma^\mu \gamma^{\rho\sigma} \gamma^\nu \gamma^{\delta\eta} c_{\rho\sigma} c_{\delta\eta}) \nonumber\\
&=& - 4 (4 a^2 + 2 b^\mu b_\mu ) \ ,\nonumber
\end{eqnarray}
where we used $[\gamma^{\rho\sigma} , \gamma^\nu] = -2 (g^{\sigma\nu} \gamma^\rho - g^{\rho\nu} \gamma^\sigma )$. Using this one easily gets the trace of the endomorphism $E$
\begin{eqnarray}\label{TrE}
\Tr E &=&  ( -12 a^2 + 8 c_{\mu\nu}c^{\mu\nu} +R)\ .
\end{eqnarray}

As a simple check of the result (\ref{TrE}) we can easily verify that the result really should not depend on $b_\mu$. If $F=\gamma^\mu b_\mu$ then the connection (\ref{Omega}) will become $\Omega_\mu = \omega_\mu + b_\mu$. This is nothing but the connection on the twisted spinor bundle and the well-known generalization of the Lichnerowicz formula \cite{Berline} immediately gives
\begin{eqnarray}
\gamma^\mu \nabla^b_\mu \gamma^\nu \nabla^b_\nu = \Delta^b +\frac{{R}}{4} + \frac{1}{2}\gamma^{\mu\nu}F_{\mu\nu} \ ,\nonumber
\end{eqnarray}
where $\nabla^b_\mu = \partial_\mu + \omega_\mu + b_\mu$ and $F_{\mu\nu} = \partial_\mu b_\nu - \partial_\nu b_\mu$ is the twisting curvature. One can easily see that this coincides with our result (\ref{E}) for this specific form of of $F$. Because the trace of the commutator is zero, the dependence on $b_\mu$ drops out of (\ref{TrE}).

\section{The relation between $\tilde{R}$ and $R$}\label{RRtil}

In this appendix we would like to obtain the relation between the Ricci scalars for a manifold endowed with two different Riemannian structures $g$ and $\tilde{g}$:
\begin{eqnarray}
{g}_{\mu \nu} = \epsilon n_\mu n_\nu + h_{\mu\nu} \nonumber \\
\tilde{g}_{\mu \nu} = \epsilon n_\mu n_\nu + \alpha h_{\mu\nu} \label{gtil} \ .
\end{eqnarray}
So, we want to relate $R\equiv R[g]$ and $\tilde{R} \equiv R[\tilde{g}]$. Define two connections, $\nabla$ and $\tilde\nabla$, compatible with $g$ and $\tilde{g}$ respectively. Then their difference is a tensor:
\begin{eqnarray}\label{delta_tensor}
(\tilde\nabla_\mu - \nabla_\mu)\omega_\nu = - C^\lambda_{\mu\nu}\omega_\lambda \ ,
\end{eqnarray}
where $\omega$ is arbitrary 1-form and
\begin{eqnarray}\label{C}
C^\lambda_{\mu\nu} = \frac{1}{2} \tilde{g}^{\lambda \sigma}(\nabla_\mu \tilde{g}_{\nu \sigma} + \nabla_\nu \tilde{g}_{\mu \sigma} - \nabla_\sigma \tilde{g}_{\mu \nu}) \ ,
\end{eqnarray}
see, e.g. Appendix D in \cite{Wald:1984rg} for details. Using (\ref{delta_tensor}) in the definition of the Riemann tensor
\begin{eqnarray}
[\nabla_\mu , \nabla_\nu]\omega_\sigma = {R_{\mu\nu\sigma}}^\rho \omega_\rho \nonumber
\end{eqnarray}
we arrive at
\begin{eqnarray}
\tensor{\tilde{R}}{_\mu_\nu_\rho^\sigma} = \tensor{R}{_\mu_\nu_\rho^\sigma} - 2 \nabla_{[\mu}\tensor{C}{^\sigma_{\nu]}_\rho} + 2 \tensor{C}{^\lambda_\rho_{[\mu}}\tensor{C}{^\sigma_{\nu]}_{\lambda}} \nonumber
\end{eqnarray}
or for the Ricci scalar
\begin{eqnarray}\label{tilR}
\tilde{R} = \tilde{g}^{\mu\nu} R_{\mu\nu} - 2 \tilde{g}^{\mu\sigma} ( \nabla_{[\mu}\tensor{C}{^\nu_{\nu]}_\sigma} - \tensor{C}{^\lambda_\sigma_{[\mu}}\tensor{C}{^\nu_{\nu]}_{\lambda}} ) \ .
\end{eqnarray}

a) Calculating $\tilde{g}^{\mu\nu} R_{\mu\nu}$.

From (\ref{gtil}) we get
\begin{eqnarray}
\tilde{g}^{\mu \nu} = \epsilon n^\mu n^\nu + \frac{1}{\alpha } h^{\mu\nu} \equiv \frac{1}{\alpha } g^{\mu\nu} + \epsilon\left(1 - \frac{1}{\alpha }\right)n^\mu n^\nu \ , \nonumber
\end{eqnarray}
where all the indices are raised by $g^{\mu\nu}$. Then we immediately get:
\begin{eqnarray}\label{R1}
\tilde{g}^{\mu\nu} R_{\mu\nu} =  \frac{1}{\alpha} R + \epsilon \bigg( 1- \frac{1}{\alpha} \bigg) R_{\mu\nu}n^\mu n^\nu \ .
\end{eqnarray}
Because from the point of view of a foliated geometry the $(3+1)$-splitting is more fundamental, we will re-write (\ref{R1}) in $3+1$ form. Using the Gauss-Codazzi equations, we get:
\begin{eqnarray}
R_{\mu\nu} n^\mu n^\nu = K^2 - K_{\mu\nu} K^{\mu\nu} + \nabla_\mu \bigg(n^\nu \nabla_\nu n^\mu - n^\mu \nabla_\nu n^\nu\bigg) \ , \nonumber \\
R =  \threeR + \epsilon K^2 - \epsilon K_{\mu\nu} K^{\mu\nu} + 2 \epsilon \nabla_\mu \bigg( n^{\lambda} \nabla_{\lambda} n^\mu - n^\mu \nabla_{\lambda} n^\lambda \bigg) \ ,\nonumber
\end{eqnarray}
where $K_{\mu\nu}$ is given by (\ref{externalcurvature}). Combining this with (\ref{R1}) we arrive at the desired result:
\begin{eqnarray}\label{R2}
\tilde{g}^{\mu\nu} R_{\mu\nu} =  \frac{1}{\alpha} \threeR + \epsilon K^2 - \epsilon K_{\mu\nu} K^{\mu\nu} + \epsilon\frac{1+\alpha}{\alpha} \nabla_\mu \bigg( n^{\lambda} \nabla_{\lambda} n^\mu - n^\mu \nabla_{\lambda} n^\lambda \bigg) \ .
\end{eqnarray}

b) Calculating $\tilde{g}^{\mu\sigma} ( \nabla_{[\mu}\tensor{C}{^\nu_{\nu]}_\sigma} - \tensor{C}{^\lambda_\sigma_{[\mu}}\tensor{C}{^\nu_{\nu]}_{\lambda}} )$.

From the metric compatibility of $\nabla$ and taking into account that $\tilde{g}_{\mu \nu}  = {\alpha} g_{\mu\nu} + \epsilon ( 1 - {\alpha} ) n_\mu n_\nu$ we have $\nabla_\mu \tilde{g}_{\nu \rho} = \epsilon ( 1 - {\alpha} )\nabla_\mu n_\nu n_\rho$. Using this in (\ref{C}) after some straightforward algebra we arrive at the following expression for $\tensor{C}{^\lambda_{\mu\nu}}$:
\begin{eqnarray}\label{C2}
\tensor{C}{^\lambda_{\mu\nu}} = \epsilon(\alpha -1)\bigg(n^\lambda K_{\mu\nu} -\epsilon \frac{1}{\alpha} n_\mu n_\nu n^\rho \nabla_\rho n^\lambda \bigg) \ .
\end{eqnarray}
From here we get
\begin{eqnarray}
\tensor{C}{^\lambda_\sigma_{[\mu}}\tensor{C}{^\nu_{\nu]}_{\lambda}} =  \frac{(\alpha -1)^2}{\alpha} K_{\nu (\sigma}n_{\mu )} n^\gamma
\nabla_\gamma n^\nu  \ .\nonumber
\end{eqnarray}
Because one of the indices $(\mu , \sigma)$ is always $3d$ and the other normal we immediately have
\begin{eqnarray}\label{CC}
\tilde{g}^{\mu\sigma} \tensor{C}{^\lambda_\sigma_{[\mu}}\tensor{C}{^\nu_{\nu]}_{\lambda}} = 0 \ .
\end{eqnarray}
After some lengthy but straightforward algebra we get for the remaining term the following result:
\begin{eqnarray}\label{nablaC}
\tilde{g}^{\mu\sigma} \nabla_{[\mu}\tensor{C}{^\nu_{\nu]}_\sigma} = \epsilon\frac{\alpha -1}{2\alpha} \nabla_\nu [n^\nu \nabla_\lambda n^\lambda + n^\lambda \nabla_\lambda n^\nu] \ .
\end{eqnarray}

Combining (\ref{tilR}), (\ref{R2}), (\ref{CC}) and (\ref{nablaC}) we arrive at the desired relation between $\tilde{R}$ and $R$
\begin{eqnarray}\label{tilR1}
\tilde{R} =  \frac{1}{\alpha} \threeR + \epsilon K^2 - \epsilon K_{\mu\nu} K^{\mu\nu} + 2 \epsilon \nabla_\mu \bigg( \frac{1}{\alpha} n^{\lambda} \nabla_{\lambda} n^\mu - n^\mu \nabla_{\lambda} n^\lambda \bigg) \ .
\end{eqnarray}

\section{The rescaled Dirac Operator}\label{rescaledDirac}

Here we would like to re-write the generalized Dirac operator (\ref{aniso-d-1}) in the form that explicitly uses the re-scaled metric (\ref{gtil}), i.e. we want to write $\mathbb{D}$ as
\begin{eqnarray}
\mathbb{D} = \tilde{D} + S \ ,
\end{eqnarray}
where $\tilde{D} = \tilde{\gamma}^\mu \nabla_\mu^{\tilde{\omega}}$ is the Dirac operator with respect to the metric (\ref{gtil}) and $S$ is some non-derivative part.

To begin with, we will re-write (\ref{aniso-d-1}) using some formulas used for the (3+1)-decomposition of a Dirac operator (see \cite{Pinzul:2014mva} for details):
\begin{eqnarray}\label{decomposition}
h^{\mu\nu} \gamma_\mu \nabla^\omega_\nu = \threeD - \frac{1}{2}\gamma^0 K\ , \nonumber \\
\epsilon n^\mu n^\nu \gamma_\mu \nabla^\omega_\nu = \gamma^0 D_n + \frac{1}{2}\gamma^\alpha\frac{\partial_\alpha N}{N}\ .
\end{eqnarray}
Using these relations in (\ref{aniso-d-1}) we get
\begin{eqnarray}\label{rescaled1}
\mathbb{D} &=& \gamma^0 D_n + c_1\threeD + c_2 \gamma^0 K + c_3 \gamma^{\alpha} a_\alpha + c_4 K + c_5 \gamma^0 \gamma^{\alpha} \frac{\partial_\alpha N}{N} = \nonumber\\
&=& (\epsilon n^\mu n^\nu + c_1 h^{\mu\nu}) \gamma_\mu \nabla^\omega_\nu + \left( \frac{c_1}{2} + c_2 \right)\gamma^0 K + \left( c_3 - \frac{1}{2} \right)\gamma^\alpha\frac{\partial_\alpha N}{N} + c_4 K + c_5 \gamma^0 \gamma^{\alpha} \frac{\partial_\alpha N}{N} \equiv \nonumber \\
&\equiv &\tilde{\gamma}^\mu \nabla_\mu^{{\omega}} + \left( \frac{c_1}{2} + c_2 \right)\gamma^0 K + \left( c_3 - \frac{1}{2} \right)\gamma^\alpha\frac{\partial_\alpha N}{N} + c_4 K + c_5 \gamma^0 \gamma^{\alpha} \frac{\partial_\alpha N}{N} \ ,
\end{eqnarray}
where we used the (3+1)-decomposition of the usual Dirac operator (\ref{D_undeformed}), which immediately follows from (\ref{decomposition})
\begin{eqnarray}
{D}= {\gamma}^\mu \nabla_\mu^{{\omega}} = \gamma^0 D_n +\threeD -  \frac{1}{2} \gamma^0 K + \frac{1}{2} \gamma^{\alpha} \frac{\partial_\alpha N}{N}	\ .\nonumber
\end{eqnarray}
and defined the re-scaled gamma matrices: $\tilde{\gamma}^\nu := (\epsilon n^\mu n^\nu + c_1 h^{\mu\nu}) \gamma_\mu$. Using the tetrads $\{ e_a^{\ \mu}\}$ adopted to the (3+1) splitting, as in the appendix \ref{notations}, it easy to see that $\tilde{\gamma}^\mu = \tilde{e}_a^{\ \mu}\gamma^a$ where $a$ is the ``flat'' index and $\tilde{e}_a^{\ \mu} = (n^\mu , c_1 {e}_i^{\ \mu})$ are the re-scaled tetrads that correspond to the metric (\ref{gtil}) with $\alpha = 1/c_1^2$. The equation (\ref{rescaled1}) is not yet what we want: while the gamma matrices are already changed to the re-scaled ones, the covariant derivative is still defined with respect to the old metric. So we need to find the relation between $\omega_\mu$ and $\tilde{\omega}_\mu$.

Using the definition of the spin connection, $\omega_\mu = -\frac{1}{4}\omega_{\mu ab}\gamma^{ab}$, where $\omega_{\mu ab} = e_{a\nu}\nabla_\mu e_b^{\ \nu} $ and the equation (\ref{delta_tensor}) we get
\begin{eqnarray}
\tilde{\omega}_{\mu ab} =  \tilde{e}_{a\nu}\tilde{\nabla}_\mu \tilde{e}_b^{\ \nu} = \tilde{e}_{a\nu}{\nabla}_\mu \tilde{e}_b^{\ \nu} + \tilde{e}_{a\nu}C^\nu_{\mu\rho} \tilde{e}_b^{\ \rho} \ .\nonumber
\end{eqnarray}

a) Calculating $\tilde{e}_{a\nu}{\nabla}_\mu \tilde{e}_b^{\ \nu}\gamma^a \gamma^b$

\begin{eqnarray}
\tilde{e}_{a\nu}{\nabla}_\mu \tilde{e}_b^{\ \nu}\gamma^a \gamma^b &=& c_1 {n}_{\nu}{\nabla}_\mu {e}_i^{\ \nu}\gamma^0 \gamma^i + \frac{1}{c_1} {e}_{i\nu}{\nabla}_\mu {n}^{\nu}\gamma^i \gamma^0 + {e}_{i\nu}{\nabla}_\mu {e}_j^{\ \nu}\gamma^i \gamma^j = \nonumber \\
&=& {e}_{a\nu}{\nabla}_\mu {e}_b^{\ \nu}\gamma^a \gamma^b + \left( c_1 + \frac{1}{c_1} -2 \right){e}_{i\nu}{\nabla}_\mu {n}^{\nu}\gamma^i \gamma^0 = \nonumber \\
&=& {e}_{a\nu}{\nabla}_\mu {e}_b^{\ \nu}\gamma^a \gamma^b + \frac{(c_1 - 1)^2}{c_1} {e}_i^{\ \nu}(K_{\mu\nu} - \epsilon n_\mu n^\sigma{\nabla}_\sigma {n}_{\nu} )\gamma^0 \gamma^i \ .\nonumber
\end{eqnarray}

b) Calculating $\tilde{e}_{a\nu}C^\nu_{\mu\rho} \tilde{e}_b^{\ \rho}\gamma^a \gamma^b$

Using (\ref{C2}) we get
\begin{eqnarray}
\tilde{e}_{a\nu}C^\nu_{\mu\rho} \tilde{e}_b^{\ \rho}\gamma^a \gamma^b &=& \epsilon \frac{1-c_1^2}{c_1^2} \tilde{e}_{a\nu}  \bigg(n^\nu K_{\mu\rho} - \epsilon c_1^2 n_\mu n_\rho n^\lambda \nabla_\lambda n^\nu \bigg)  \tilde{e}_b^{\ \rho}\gamma^a \gamma^b = \nonumber \\
&=& \epsilon \frac{1-c_1^2}{c_1^2} \left( \epsilon c_1 K_{\mu\nu} {e}_i^{\ \nu}\gamma^0 \gamma^i - \frac{c_1^2}{c_1} n_\mu e_{i\nu} n^\lambda \nabla_\lambda n^\nu \gamma^i \gamma^0 \right) = \nonumber \\
&=& \frac{1-c_1^2}{c_1} \left( K_{\mu\nu} {e}_i^{\ \nu} + \epsilon n_\mu e_{i\nu} n^\lambda \nabla_\lambda n^\nu \right) \gamma^0 \gamma^i \ .\nonumber
\end{eqnarray}

Combining (a) and (b) we get
\begin{eqnarray}\label{omegatil}
\tilde{e}_{a\nu}\tilde{\nabla}_\mu \tilde{e}_b^{\ \nu} \gamma^a \gamma^b &=& {e}_{a\nu}{\nabla}_\mu {e}_b^{\ \nu}\gamma^a \gamma^b + 2 \frac{1-c_1}{c_1} {e}_i^{\ \nu} K_{\mu\nu} \gamma^0 \gamma^i -2\epsilon (c_1 - 1)n_\mu e_{i\nu} n^\lambda \nabla_\lambda n^\nu \gamma^0 \gamma^i = \nonumber \\
&=& {e}_{a\nu}{\nabla}_\mu {e}_b^{\ \nu}\gamma^a \gamma^b + 2 \frac{1-c_1}{c_1} {e}_i^{\ \nu} K_{\mu\nu} \gamma^0 \gamma^i + 2(c_1 - 1)n_\mu e_{i\nu} \frac{\partial_\nu N}{N}\gamma^0 \gamma^i \ ,
\end{eqnarray}
where we used the identity (B.7) from \cite{Pinzul:2014mva}:
\begin{eqnarray}
n^\mu n_\nu\nabla_\mu e_{i}^{\ \nu}\gamma^0  \gamma^i = \epsilon e_{i}^{\ \alpha}\frac{\partial_\alpha N}{N} \gamma^0 \gamma^i \ .\nonumber
\end{eqnarray}

Using (\ref{omegatil}) we obtain
\begin{eqnarray}
\tilde{\gamma}^\mu \omega_\mu &=& \tilde{\gamma}^\mu \tilde{\omega}_\mu + \frac{1-c_1}{2 c_1} \tilde{\gamma}^\mu {e}_i^{\ \nu} K_{\mu\nu} \gamma^0 \gamma^i + \frac{c_1 - 1}{2}\tilde{\gamma}^\mu n_\mu e_{i\nu} \frac{\partial_\nu N}{N}\gamma^0 \gamma^i = \nonumber \\
&=& \tilde{\gamma}^\mu \tilde{\omega}_\mu + \frac{1-c_1}{2}\left( \gamma^0 K + \gamma^i e_i^{\ \mu}\frac{\partial_\mu N}{N} \right) \ .\nonumber
\end{eqnarray}
Using this result in (\ref{rescaled1}) we finally arrive at the desired relation (\ref{rescaled2})
\begin{eqnarray}\label{rescaled3}
\mathds{D} = \tilde{\gamma}^\mu \nabla_\mu^{\tilde{\omega}} + \bigg( c_2 + \frac{1}{2} \bigg) K \gamma^0 + \bigg(c_3 - \frac{c_1}{2} \bigg) \gamma^i e_i^{\ \mu}\frac{\partial_\mu N}{N} + c_4 K +  c_5 \gamma^0 \gamma^i e_i^{\ \mu}\frac{\partial_\mu N}{N} \ .
\end{eqnarray}

\section{3+1 Metric with spherical symmetry}\label{ap:spherical}

The most general three-dimensional spherically symmetric metric can be written in the coordinates adopted to the spherical symmetry as
\begin{equation}
{}^{(3)}g_{ij} = \text{diag} \left(a^2(r,t), r^2 b^2(r,t), r^2 b^2(r,t) \sin^2(\theta)\right).\nonumber
\end{equation}
The spherical symmetry constrains the shift vector $N^i$ in the ADM decomposition
\begin{equation}
ds^2  = \alpha^2 dt^2 + {}^{(3)}g_{ij} (dx^i + N^i dt)(dx^j + N^j dt),\nonumber
\end{equation}
to
\begin{equation}
N^i = (n(r,t), 0, 0).\nonumber
\end{equation}
Thus, the ADM decomposition of a spherically symmetrical metric can be written as
\begin{equation}
ds^2  = (N^2 + a^2 n^2)dt^2 + 2 a^2 n dt dr + a^2 dr^2 + r^2 b^2 d\Omega^2.\nonumber
\end{equation}

In \cite{Barausse:2011pu,Barausse:2013nwa} it was obtained a solution to a black-hole in
Aether-Einstein theory, that is also a solution to the low energy limit of Horava-Lifshitz gravity,
\begin{equation}
ds^2 =  f(r) dv^2 + 2 B(r) dv dr + r^2 d\Omega^2 \ ,\nonumber
\end{equation}
where $v$ is the Eddington-Finkelstein coordinate $v = t + r^\ast$, with $r^\ast$
such that $\frac{dr^\ast}{dr}=B/f$, and
\begin{eqnarray}
f(r)&=& \left( 1+\frac{F_1}{r} + \frac{\eta}{48\xi} \frac{F_1^3}{r^3} + \ldots \right) \nonumber \\
B(r)&=& \left( 1 + \frac{\eta}{16\xi} \frac{F_1^2}{r^2} + \frac{\eta}{12\xi} \frac{F_1^3}{r^3} + \ldots \right) \ , \nonumber
\end{eqnarray}
where $F_1$  is proportional to the total mass $M$ of a black hole as seen by a distant observer $F_1 = - 2 G_N M$ and $\eta$ and $\xi$ are the parameters that enter the IR action of HL gravity (\ref{IRHL}).

Switching back to Schwarzschild coordinates

\begin{equation}
ds^2 = f(r) dt^2 + \frac{B^2}{f} dr^2 + r^2 d\Omega^2 \ ,\nonumber
\end{equation}
we can readily identify
\begin{equation}
N(r) = \sqrt{f(r)}\  , \quad  a(r) = \frac{B(r)}{\sqrt{f(r)}}\ , \quad n = 0\ .\nonumber
\end{equation}

This solution implies that Earth-based experiments in our scenario would provide a contribution to the radial component of the Kostelecky parameter $m_\mu$ in \eqref{LVT} given by

\begin{equation}\label{m_r}
m_r = \left( c_3 - \frac{1}{2} \right) a(r) = \left( c_3 - \frac{1}{2}\right) \frac{G_N M}{ r^2} \left( 1 + \frac{G_N^2 M^2}{4 r^2}  \frac{\eta}{\xi} + \ldots \right) .
\end{equation}

\end{appendices}

\bibliographystyle{utphys}
\bibliography{actiondiracbib}

\end{document}